\documentclass[12pt]{iopart}
\usepackage{amssymb}
\usepackage{iopams}
\usepackage{amsthm}
\usepackage{latexsym}
\usepackage{hyperref}
\usepackage{fontenc}
\usepackage{geometry}
\usepackage{setspace}

\newcommand{\bdm}{\begin{displaymath}}
\newcommand{\edm}{\end{displaymath}}
\newcommand{\be}{\begin{equation}}
\newcommand{\ee}{\end{equation}}
\newcommand{\bea}{\begin{eqnarray}}
\newcommand{\nn}{\nonumber}
\newcommand{\eea}{\end{eqnarray}}

\begin{document}

\title[L. A. Pach\'{o}n and J. D. Sanabria-G\'{o}mez]
      {On the reflection symmetry in stationary axisymmetric
       electrovacuum space-times}

\author{
        Leonardo A. Pach\'{o}n$^{1,2}$
                \footnote[1]{e-mail: lpachon@laft.org} and
        Jos\'e D. Sanabria-G\'{o}mez$^{2}$
                \footnote[2]{e-mail: jsanabri@uis.edu.co}}

\address{$^1$
            Laboratorio de Astronom\'ia y F\'isica Te\'orica (LAFT),
Departamento de F\'isica, Facultad de Ciencias, La Universidad del
Zulia, Maracaibo, 4004, Venezuela. \\
         $^2$
              Escuela de F\'isica, Universidad Industrial de Santander. A.A.
              678, Bucaramanga, Colombia.
              }

\begin{abstract}
Recently Kordas (1995 {\it Class. Quantum Grav.} \textbf{12} 2037) and Meinel
and Neugebauer (1995 {\it Class. Quantum Grav.} \textbf{12} 2045) studied the
conditions for reflection  symmetry  in stationary axisymmetric space--times in
vacuum. They found that a solution to the Einstein field equations is
reflectionally symmetric if   their  Ernst's potential  ${\cal E}(\rho=0,z)=
e(z)$  on a portion of the positive z-axis extending to  infinity  satisfies
the condition $e_{+}(z){e}_{+}^{*}(-z)=1$. In this note, we formulate an
analogous conditions for two complex Ernst potentials in electrovacuum.
We also present the special case of rational axis potentials.

\end{abstract}

\vspace{1cm}

It is well known that a metric manifold $({\cal{M,}}  h)$ is asymptotically
flat  if  another  manifold exists, {\it i. e.} $(\tilde{\cal{M}},\tilde{h})$
with $\tilde{\cal{M}}= {\cal{M}}\cup\Lambda$, and their metrics are
related by a transformation of the form $h_{ij} \rightarrow \tilde{h}_{ij} =
\Omega^2h_{ij},$ where $\Lambda$ is the point added to the initial manifold
to represent infinity. In case of a stationary axisymmetry, the $h$ metric
of manifold $\cal M$ in Weyl-Papapetrou's coordinates has the form

\be
    \label{metric}
    ds^2= -F(dt-\omega d\phi)^2+F^{-1}\left[e^{2\gamma}(d\rho^2+dz^2)+
    \rho^2 d\phi^2 \right]\, ,
\ee where $F,\omega$, and $\gamma$ are functions of $\rho$ and
$z$ only. We shall analyze the reflection condition by using the Ernst's
potentials $\cal E$ and $\Phi$ \cite{Ernst} defined as

\be
    {\cal{E}}= F+i\Psi-\Phi\Phi^{*}\, ,
    \qquad
    \Phi=A_{4}+iA_{3}^{'}\, ,
\ee
where $*$ denotes usual complex conjugation. $A^{'}_3$  and $\Psi$ satisfy

\begin{eqnarray}
\label{A3A4}
    \rho^{-1} f (\nabla A_4 - \omega \nabla A_3)= \hat{e}_\phi \times \nabla
    A^{'}_3\, ,
\\
    \nabla \cdot \{f^{-2}[\nabla \Psi+2\rm{Im}(\Phi^* \nabla
    \Phi)]\}=0\,,
\end{eqnarray}
where $A_4$ and $A_3$ denote the time and azimuthal component of the electromagnetic four-potential
$A_{\alpha}$, and $\nabla$ is the usual three dimensional divergence operator in Weyl coordinates.
Using the Ernst's potentials $\cal E$ and  $\Phi$ the Einstein-Maxwell's equations becomes

\begin{eqnarray}
     \label{eFe}
  F&=& {\rm Re}\,{\cal E}+ \Phi \Phi^*,\\
    \label{omega}
\nabla\omega&=& \rho F^{-2} [\hat{e}_\phi \times \nabla \Psi+
            2\hat{e}_\phi \times {\rm Im}(\Phi^* \nabla \Phi)],
\\
 \label{gamma}
        \gamma_{,\rho} &=& \frac{\rho}{4({\rm Re}\,{\cal E}
        +\Phi \Phi^*)^2}[({\cal E}_{,\rho}+2 \Phi^*
        \Phi_{,\rho})({\cal E}^*_{,\rho}+2\Phi \Phi^*_{,\rho})-
        \nonumber \\ &-&({\cal E}_{,z}+2\Phi^* \Phi_{,z}) ({\cal
        E}^*_{,z}+2\Phi \Phi^*_{,z})]-\frac{\rho(\Phi_{,\rho}
        \Phi^*_{,\rho}-\Phi_{,z}\Phi^*_{,z})}{{\rm Re}\,{\cal E}+\Phi\Phi^*}, \nonumber
    \\
 \gamma_{,z} &=& \frac{\rho {\rm Re}[({\cal
        E}_{,\rho} +2\Phi \Phi^*_{,\rho})({\cal E}^*_{,z} + 2\Phi \Phi^*_{,z})]} {2({\rm
        Re}\,{\cal E}+\Phi \Phi^*)^2} -\frac{2\rho {\rm Re} (\Phi^*_{,\rho} \Phi_{,z})}
        {{\rm Re}\,{\cal E} +\Phi \Phi^*}\, .
        \\ \nn
\end{eqnarray}
Physically,  when  a  manifold $\cal M$ has an orthogonal symmetry
plane  to the $z$-axis, the multipolar moments of even order are
real and  those  of  odd  order  are purely imaginary.  Because a multipole structure  defines  a  space--time  uniquely,  it is
possible to formulate  a  necessary and sufficient condition for
a space--time to be reflectionally symmetric in terms of
multipole moments. In that sense,  we  introduce  the complex
Ernst's potentials $\xi$ and $q$, which  are  the analogous of
Newtonian gravitational potential and Coulomb potential
respectively,

\bea
    \label{Efunction}
    {\cal E}=\frac{1-\xi}{1+\xi}\, ,
\qquad
    \Phi=\frac{q}{1+\xi}.
\eea Since  the  multipolar  structure  of  a  solution is
calculated at infinity  \cite{HoensPerj},  the $\tilde h$ metric is
used  and  the  coordinate  transformations $ \bar{\rho}=\rho/ (\rho
^2  + z^2),$ $\bar {z}=z/(\rho ^2 + z^2)$ and $\bar  {\phi  }=
\phi$ are introduced to bring infinity at the origin of the axes.
Choosing the conformal factor $\Omega=
\bar{r}^2=\bar{\rho}^2+\bar{z}^2\,  ,$ the $\xi$ and $q$ potentials
transform as

\be
    \tilde{\xi}=\Omega^{-1/2} \xi ,\qquad \tilde{q}=\Omega^{-1/2}q.
\ee
Because of the axisymmetry,  the  components  of  the  tensorial moments
are  multiples  of  the corresponding scalar moments, which are the
projection  of  the  tensorial  moments  on  the  axis of symmetry
\cite{HoensPerj}.   The  scalar  moments  are  calculated  as  the
coefficients  in power series of $\tilde  \xi$ and $\tilde q$ in terms of
$\bar{\rho}$ and $\bar{z}$. By using the recursive
expressions   (20)   and   (21)   from  \cite{Apostolatos},  these
coefficients  are  determined  by  their form on the symmetry axis, therefore

\be
\label{multipolesaxis}
    \tilde{\xi}(\bar\rho=0)=\sum_{i=0}^{\infty} m_i
    {\bar z}^i ,\qquad \tilde{q}(\bar\rho=0)=\sum_{i=0}^{\infty} q_i
    {\bar z}^i.
\ee
Using these findings and the fact that if the multipole moments are real
(imaginary) on the symmetry axis then they are also real (imaginary) in the whole
space (see \cite{Kordas} for a proof of this statement in the vacuum case, the
electrovacuum generalisation is straightforward), we assume that
our analysis on the symmetry axis is valid for the whole space--time, therefore
we need to focus exclusively on the symmetry axis.

\vspace{0.25cm}

It   is   easy   to   verify  in  (\ref{eFe}),  (\ref{omega})  and
(\ref{gamma})  that  if $A_{3}$ and $A_{4}$ are either both odd or
both  even  functions  of  $z$  then  it  is  consistent  with the
reflection symmetry of the metric functions. We analyze
these two  cases assuming that they are the only physically relevant
ones \cite{ASotiriou}, and we derive conditions for the Ernst
potentials on the symmetry axis. Finally, in order to obtain reflectionally symmetric
solutions we derive general expressions for these conditions.

\section*{First case. $A_{3}$ and $A_{4}$ are both even functions of
$z$.}

To obtain the conditions for Ernst potentials in the
reflectionally symmetry case, we assume that the metric is
also reflectionally symmetric
and derive these conditions from field equations. From
(\ref{A3A4}), (\ref{eFe}) and (\ref{omega}) it is clear that if
$F$, $\omega$, $A_{3}$ and $A_{4}$ are reflectionally symmetric
the Ernst's potentials satifies ${\cal {E}}(\rho, -z)={\cal{E}^{*}} (\rho,  z)$ and $\Phi
(\rho, -z)=\Phi^{*}(\rho, z)$ and, from (\ref{gamma}), $\gamma$
is also reflectionally symmetric.
Using these conditions in (\ref{Efunction}) we have that ${\xi}(\rho,-z)
={\xi^{*}}(\rho,z)$ and $q(\rho, -z)=q^{*}(\rho, z)$; moreover
$\tilde {\xi}$    and $\tilde{q}$ satisfy ${\tilde {\xi}}(\rho,
-z)={\tilde{\xi}^{*}}(\rho,    z) $ and ${\tilde{q}}(\rho,
-z)={\tilde{q}^{*}}(\rho,  z)$.

\vspace{0.25cm}

When these conditions are applied to the symmetry axis, the metric functions
are reflectionally symmetric if the Ernst's potentials obey the functional
relations \be e_{+}(z){e}_{+}^{*}(-z)=1\, , \qquad
f_{+}(z)=-{f}_{+}^{*}(-z)e_{+}(z)\, , \label{condef1} \ee
for a portion  of  the  positive $z$-axis
extending to infinity, $e_{+}$ and $f_{+}$. This guarantees that
the operation ${z}\rightarrow {-z}$ (or
equivalently ${\bar z}\rightarrow {-\bar z}$) has the effect of
${\tilde{\xi}}\rightarrow {\tilde {\xi}}$ and ${\tilde
{q}}\rightarrow {\tilde {q}}$. Hence the coefficients of
${\tilde{\bar\xi}}$ and ${\tilde {q}}$ satisfy the reflection
symmetry  condition in a neighbourhood of $\Lambda$ and the
positive part of the axis if the Ernst's potentials satisfy
(\ref{condef1}).

\vspace{0.25cm}

When the Ernst's potentials of the metric $h$  can be
written as rational functions in Weyl coordinates,
for a portion
of the positive z-axis extending to infinity, $e_{+}$ and $f_{+}$, they
will have the form

\be
    {\footnotesize e(z)=\frac{z^{N}+{
    \sum_{j=1}^{N}(-1)^{j}\Upsilon_{j}z^{N-j}}}{z^{N}+{
    \sum_{j=1}^{N}{\Upsilon}_{j}^{*}z^{N-j}}}}\, ,
\quad
    {\footnotesize f(z)=\frac{ \sum_{j=1}^{N}(-i)^{j+1}
    \Lambda_{j}z^{N-j}} {z^{N} +
    {\sum_{j=1}^{N}\Upsilon_{j}^{*}z^{N-j}}}},
\label{e(z)f(z)1}
\ee
with $\Upsilon_{k}=D_{k}+iS_{k},\,$ and
$D_{k},\, S_{k},\, \Lambda_{k}\,\in\mathbb{R}.$

\section*{Second case. $A_{3}$ and $A_{4}$ are both odd
functions of $z$.}

Following a similar procedure similar to the one used in the previous case,
it is possible to show that ${\tilde {\xi}}(\rho,
-z)={\tilde{\xi}^{*}}(\rho, z) $ and ${\tilde{q}}(\rho,
-z)=-{\tilde{q}^{*}}(\rho, z)$. The reflection symmetry conditions
for this case are given by

\be
    e_{+}(z){e}_{+}^{*}(-z)=1\, , \qquad
    f_{+}(z)={f}_{+}^{*}(-z)e_{+}(z)\, . \label{condef2}
\ee
Anologuos to  (\ref{e(z)f(z)1}),   if  the Ernst's potentials can be written as rational functions
in  Weyl  coordinates,  then  they could be written as

\be
    {\footnotesize e(z)=\frac{z^{N}+{\sum_{j=1}^{N}(-1)^{j}
    \Upsilon_{j}z^{N-j}}}{z^{N}+{
    \sum_{j=1}^{N}{\Upsilon}_{j}^{*}z^{N-j}}}}\, ,\quad
    {\footnotesize f(z)=\frac{ \sum_{j=1}^{N}(-i)^{j+2}
    \Lambda_{j}z^{N-j}} {z^{N} +
    {\sum_{j=1}^{N}\Upsilon_{j}^{*}z^{N-j}}}},
\label{e(z)f(z)2} \ee with $\Upsilon_{k}=D_{k}+iS_{k},\,$ and
$D_{k},\, S_{k},\, \Lambda_{k}\,\in\mathbb{R}.$
\vspace{0.25cm}

This second case  can be obtained by the analytic
continuation $\Lambda \rightarrow  i  \Lambda$ in  (\ref{e(z)f(z)1}), which
means that our initial reflection  symmetry  condition,  {\it i. e.} multipolar
moments  of  even (odd)  order  are  real (imaginary), is not satisfied for
electromagnetic multipole moments. Instead, it satisfies the fact that
multipolar moments  of  even (odd)  order  are imaginary (real). Monopole
magnetic and dipole electric fields are examples of this second case.

\vspace{0.25cm}

Conditions  shown in (\ref{condef1}) and (\ref{condef2}) represent the
relativistic  generalisation  of the classical fact that solutions
to   the  Laplace  equation  vanishing  at  infinity  and  having
reflection   symmetry   with   respect  to  the  plane  $z=0$  are
characterised by a potential which is an odd function of $z$ on
the upper part of the $z=0$ axis. An additional study of the
structure of  the  Ernst's  potential  (\ref{e(z)f(z)1}) and
(\ref{e(z)f(z)2}) could  show that the parameters $D_{k},\,
S_{k},$ and $\Lambda_{k}$ admit a clear physical interpretation.
On the other hand, since it is  conjectured,   all  the  relevant
astrophysics  objects  are reflectionally symmetry  and their
gravitational field should be studied by  means  of  analytical
exact solutions to the Einstein-Maxwell's equations.

Thus, explicit  relations  between  the parameters of
the  Ernst's  potentials  and  the  coefficients $m_i$ and $q_i$
(Eq. \ref{multipolesaxis}), could provide a useful way to built
exact solutions which describe astrophysical objects, such as
galaxies or neutron stars. We use a more suitable
representation of $\Lambda_j$ parameters, {\it i.e.}
$\Lambda_{2j+1}=(-i)^{-j-1}V_{2j+1}$ and
$\Lambda_{2j}=(-i)^{-j}E_{2j}\,$ compared to those in (\ref{e(z)f(z)1}) to obtain
the following explicit relations

\bea
    m_{2n}&=& -\sum^{n-1}_{l=0}m'_{2n-2l-1}S_{2l+1}-
    \sum^{n}_{r=1}m_{2n-2r}D_{2r}+D_{2n+1}
\label{recumasa}
\\
    m'_{2n+1}&=&\sum^{n}_{l=0}m_{2n-2l}S_{2l+1}-
    \sum^{n-1}_{r=0}m'_{2n-2r-1}D_{2r+2}-S_{2n+2}
\label{recurot}
\\
    q_{2n}&=&-\sum^{n-1}_{l=0}q'_{2n-2l-1}S_{2l+1}-\sum^{n}_{r=1}
    q_{2n-2r}D_{2r}+V_{2n+1}
\label{recucarg}
\\
    q'_{2n+1}&=&\sum^{n}_{l=0}q_{2n-2l}S_{2l+1}-
    \sum^{n-1}_{r=0}q'_{2n-2r-1}D_{2r+2}-E_{2n+2}
\label{recumag}\, , \eea where $m'_{2n+1}$ and $q'_{2n+1}$ are the
imaginary part of $m_{2n+1}$ and $q_{2n+1}$ respectively. Notice
that the imaginary part of $m_{2n}$ and $q_{2n}$ and the real part
of $m_{2n+1}$ and $q_{2n+1}$ vanish by reflection symmetry
condition. The parameters $D_j$ and $S_j$ are related with the
mass and the angular moment multipoles and $V_j$ and $E_j$ with
the electric and magnetic ones. The sets of expressions
(\ref{recumasa})-(\ref{recumag}) and expressions (24) and (25)
from \cite{Apostolatos} allow us to calculate the higher order
multipoles moments as function of the lower ones It is also
possible to derive analogous expressions for for the Ernst's
potentials (\ref{e(z)f(z)2}).

\section*{Acknowledgments}
L. A. Pach\'{o}n  thanks Professor  Malcolm MacCallum  for
motivating the preparation of this manuscript and Doctor
Panagiotis Kordas for some suggestions. J. D. Sanabria-G\'{o}mez
acknowledges financial support from COLCIENCIAS -- Colombia and
from Project 5116 (DIF de Ciencias of the Universidad Industrial
de Santander, Colombia). The authors thank an anonymous referee
for helpful comments.

\end{document}